\documentclass[twocolumn,secnumarabic,amssymb, nobibnotes, aps, pra, groupedaddress]{revtex4-2}
\usepackage{graphicx}
\usepackage{dcolumn}
\usepackage{bm}
\usepackage[colorlinks,linkcolor=blue,anchorcolor=blue,citecolor=blue,urlcolor=blue]{hyperref}
\usepackage{mathrsfs}
\usepackage{amsmath}
\usepackage{mathtools}

\begin{document}
\title{Hybrid quantum system with strong magnetic coupling of a magnetic vortex to a nanomechanical resonator}

\author{Bo-Long Wang}
\author{Xin-Lei Hei}
\author{Xing-Liang Dong}
\author{Xiao-Yu Yao}
\author{Jia-Qiang Chen}
\author{Yi-Fan Qiao}
\author{Fu-Li Li}
\author{Peng-Bo Li}
\email{lipengbo@mail.xjtu.edu.cn}
\affiliation{Ministry of Education Key Laboratory for Nonequilibrium Synthesis and Modulation of Condensed Matter, Shaanxi Province Key Laboratory of Quantum Information and Quantum Optoelectronic Devices, School of Physics, Xi’an Jiaotong University, Xi’an 710049, China}

\begin{abstract}
We present a hybrid quantum system composed of a magnetic vortex and a nanomechanical resonator. We show that the gyrotropic mode of the vortex can coherently couple to the quantized mechanical motion of the resonator through magnetic interaction. Benefiting from  the topologically protected properties and the low damping of vortices, as well as the excellent coherent features of  nanomechanical resonators, the proposed system can achieve strong coupling and even the ultrastrong coupling regime by choosing appropriate parameters. In combination with other quantum systems, such as a nitrogen-vacancy (NV) center, coherent state transfer between the vortex excitation and the spin can be realized. This setup provides a potential platform for quantum information processing and investigations into the ultrastrong coupling regimes and macroscopic quantum physics.
\end{abstract}
\maketitle
\section{Introduction}
\label{intro}
In recent years, quantum information processing has attracted considerable attention, and the development of a quantum system platform that can achieve a strong coupling regime is at the heart of the investigation. Hybrid quantum systems, as one of the most promising candidates, have some unique superiorities that the individual component systems cannot provide, because they can integrate the advantages of different physical subsystems \cite{RevModPhys.85.623,Kurizki3866}.
Typical systems include ultracold atoms coupled to photons \cite{PhysRevLett.103.043603} and phonons \cite{j.crhy.2011.04.015}, hybrid systems based on magnonics \cite{science.aaa3693,Rab248d,PhysRevLett.113.083603,PhysRevLett.124.093602}, spins coupled to photons \cite{zhou2021chiral,PhysRevLett.112.213602}, phonons \cite{doi:10.1021/nl300775c,PhysRevA.103.013709,Kolkowitz1603,Nphys2070,PhysRevLett.117.015502,PhysRevB.79.041302,PhysRevLett.99.140403,s11128-020-02714-5,PhysRevA.100.043825}, acoustics devices \cite{PhysRevLett.126.203601,li2018hybrid,PhysRevA.101.042313,PhysRevResearch.2.013121,PhysRevResearch.3.013025,qute.202100074,PhysRevApplied.11.044026} and superconducting qubits \cite{PhysRevLett.107.220501}, photon-nonlinear medium \cite{PhysRevLett.118.223605,OE.26.011147,PhysRevLett.124.160501}, superfluid-optomechanical system \cite{arxiv.2208.05660,OE.397478,s41567-020-0785-0}, all of which fully exploit the advantages of each subsystem.
For instance, solid-state spin systems \cite{j.physrep.2013.02.001}, as well as atom-based systems possess high coherence qualities, allowing them to function as quantum memory registers. When considering interfacing various components of a hybrid system and transferring quantum information, nanomechanical resonators are great alternatives \cite{nnano.2006.208,j.physrep.2011.12.004,Kolkowitz1603} because the mechanical oscillation can achieve multifarious interactions with various types of quantum systems \cite{Nphys2070,j.crhy.2011.04.015,PhysRevLett.117.015502,PhysRevB.79.041302,PhysRevLett.99.140403}. Furthermore, by introducing a linear resource, these coupling strengths can be further enhanced \cite{PhysRevLett.125.153602}. The acoustics devices, which propagate phonons at low speed, provide special benefits for the transmission of quantum information.

Magnons, the quasiparticles of the collective spin excitations, have attracted many interests and are considered to be a promising platform for studying novel quantum technologies and macroscopic quantum phenomena \cite{science.aaa3693,Rab248d,PhysRevLett.113.083603,PhysRevLett.124.093602,PhysRevLett.121.203601,sciadv.1501286,PhysRevLett.113.156401,PhysRevLett.123.107702,PRXQuantum.2.040344}. It has been reported that magnons can interact coherently with photons in the microwave cavity and phonons via magnetostrictive interactions \cite{PhysRevLett.121.203601}. Most of the previous works are based on the uniform Kittel mode in small spheres \cite{Rab248d,PhysRevLett.121.203601,PhysRevLett.124.093602,PhysRevB.101.125404,PhysRevA.103.043706}, while few works focus on coupling photons to magnetic textures, such as vortices and skyrmions \cite{j.physrep.2020.12.004}.

A vortex \cite{PhysRevLett.83.1042,1.2221904,jnn.2008.003,s0304-8853(02)01471-3,PhysRevB.76.224426,PhysRevB.72.024455,nphys619,nnano.2016.63}, known as a topological defect, is a curling vortex structure of spin waves in thin-film confined geometries of ferromagnetic materials \cite{PhysRevB.103.064408,PhysRevB.67.020403,science.1075302}, and the stability of which can be traced to topological considerations \cite{00018732.2012.663070}. This magnetic texture originates from the competition between magnetostatic and exchange energies and is determined by the geometric dimensions and the intrinsic material properties. An external magnetic field can also affect the properties of a vortex. As a topologically protected particle-like soliton, the vortex can be characterized by the polarity ($P$) and circulation ($C$), which indicate the upward ($P=+1$) or downward ($P=-1$) orientation of the vortex core magnetization, and the clockwise ($C=+1$) or counterclockwise ($C=-1$) rotation of the in-plane magnetization, respectively. The polarity and the circulation are independent of each other, leading to a stable four-state logic unit. Furthermore, due to the existence of the energy barrier when switching states, the vortex is robust to thermal fluctuations, while the states may be changed quickly by versatile methods \cite{PhysRevLett.100.027203,nphys1810,nature05240,1.4847375,nmat1867}. This bistable property \cite{PhysRevLett.102.177602} of the vortex makes it topologically protected, and becomes a potential candidate for information storage applications \cite{1.2998584,PhysRevLett.100.027203,1.3373833}.
When stimulated by a magnetic field pulse, the core of the vortex precesses around its equilibrium position, known as the gyrotropic mode \cite{science.1095068,PhysRevLett.96.067205,1.4878617,ncomms1006,jnn.2008.003,1.1450816,PhysRevLett.100.027203}. This vortex state excitation, with a sub-gigahertz range of tunable frequency \cite{1.3012380,1.3373833,1.3563561} and narrow linewidth, can couple to the photons in a cavity \cite{PhysRevB.98.241406,acsphotonics.8b00954,ab52d7,s11128-021-03305-8}.

When the interaction strengths increase to a comparable fraction or even exceed the bare frequencies of the uncoupled subsystems, the so-called ultrastrong coupling (USC) regime arises, which has been investigated in several quantum systems \cite{RevModPhys.91.025005,s42254-018-0006-2,Nphys1730,1.4939134}. In this regime, the Jaynes-Cummings (JC) model breaks down since the rotating-wave approximation (RWA) is invalid while the intricacy of the quantum Rabi model emerges, and the system's ground state is a squeezed vacuum that contains correlated pairs of virtual excitations \cite{PhysRevB.72.115303}. Employing the unconventional natures of the USC, novel applications in quantum technologies are inspired, such as ultrafast two-qubit gates \cite{PhysRevLett.108.120501}, quantum error correction \cite{PhysRevB.91.064503}, and quantum simulations \cite{s41467-017-01061-x,s41467-017-00894-w}.

In this work, we propose a hybrid quantum device interfacing a magnetic vortex with a nanomechanical resonator. The resonator tip carries a ferromagnetic rod and creates a field with strong magnetic field gradients \cite{PhysRevLett.99.140403}, which can lead to a dynamical distortion of the vortex core in a ferromagnetic nanodisc. Thus, a vortex gyration-phonon interaction emerges and enables the exchange of quantum information between them.
The magnitude of coupling strength is relevant to the size and material properties of the vortex disc, as well as the magnetic field gradient. As the distance decreases, the coupling strength increases significantly and can reach the USC regime.
Together with a single NV center, we present an application that can equate the system to an indirect coupling between the NV center and the vortex gyration under large detuning conditions, where the phonon is only virtually excited.
Owing to the topologically protected property of the vortex and the wide application of mechanical resonators, this proposal could provide a promising platform for quantum information processing and studying quantum physics on the macroscopic scale \cite{RevModPhys.85.623} for example, the nonvolatile memory \cite{1.3373833,1.2998584,ncomms11584}, quantum entanglement \cite{PhysRevA.71.032317,s11128-021-03305-8}, spin-wave emitter \cite{nnano.2016.117}, and quantum computation \cite{nphys3347,ncomms5700}. Besides, this setup may provide a novel hybrid system with higher normalized coupling (a dimensionless parameter defined as the ratio between the coupling strength and the bare frequency of the excitations) compared with the previous work, where the coupling strength between the magnon of a sphere and the photon constitutes about $10\%$ of the photon energy \cite{PhysRevApplied.2.054002}.

\section{The setup}\label{sec:2}
\begin{figure}[htbp]
\centering\includegraphics[width=0.45\textwidth]{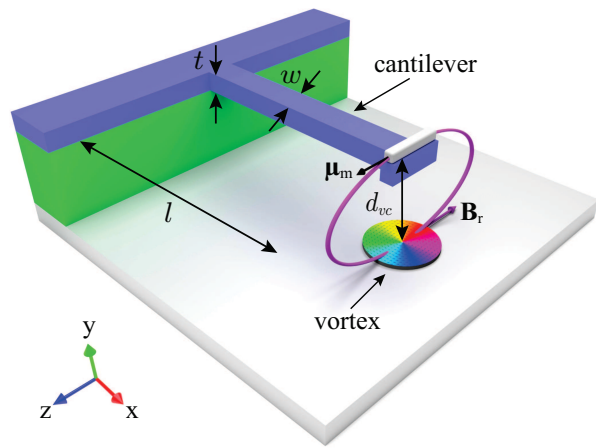}
\caption{Sketch of the proposed hybrid system. A ferromagnetic nanodisc is under the cantilever. On the free end of the cantilever, a rod-shaped single-domain ferromagnet is attached, which creates an oscillating magnetic field with a strong gradient at the center of the nanodisc. Here we choose the sizes of the cantilever ($l_{c}, w_{c}, t_{c}$) and the magnet ($l_{m}, w_{m}, t_{m}$) as: $(l_{c}, w_{c}, t_{c})=(1.2, 0.2, 0.15)$ $\mu$m, and $(l_{m}, w_{m}, t_{m})=(0.3, 0.05, 0.05)$ $\mu$m. The distance is $d_{vc}=150$ nm.
}\label{fig:1}
\end{figure}

The hybrid system under investigation is illustrated in Fig.~\ref{fig:1}. We consider a ferromagnetic nanodisc magnetically coupled to the motion of a cantilever resonator via a single domain ferromagnet attached to its end, which creates a magnetic field with strong gradients $G_{v}$.
The magnet is positioned directly above the nanodisc at a distance $d_{vc}$ and transduces the mechanical oscillation of the cantilever tip into a time-varying magnetic field.
By choosing the appropriate orientation of the magnetic moment, the core of the nanodisc is subject to a magnetic field $\mathbf{B}_{r}$ along the in-plane orientation of the disc, which can couple to the gyrotropic motion of a vortex.

The magnetic vortex is a topological structure characterized by the out-of-plane magnetization at a small region of the vortex core and the in-plane flux-closure magnetization around the core.
It has already been demonstrated that the gyrotropic mode, which corresponds to a rotation of the pattern center at a characteristic frequency in the sub-gigahertz range, is the lowest excitation mode of the vortex state, i.e., the core shifts from its equilibrium position as excited by a lateral magnetic field or current pulses and oscillates along spiral trajectories when the driving source is turned off. The value of the gyration frequency $\omega_{v}$ depends upon the aspect ratio $\beta=t/r$ of the disc, with $t$ and $r$ being the thickness and radius of the disc, respectively. For a thin disc (with aspect ratio $\beta\ll 1$) made of a certain material, the frequency is approximately proportional to $\omega_{v} \propto M_s\beta$, where $M_s$ is the saturation magnetization of the chosen material.
In order to couple this mode to the cantilever resonator, nanodiscs with small aspect ratios made out of soft ferromagnetic materials with relatively low saturation magnetization will be appropriate, and low Gilbert damping parameters facilitate the coupling.

\begin{figure}[htbp]
\centering\includegraphics[width=0.45\textwidth]{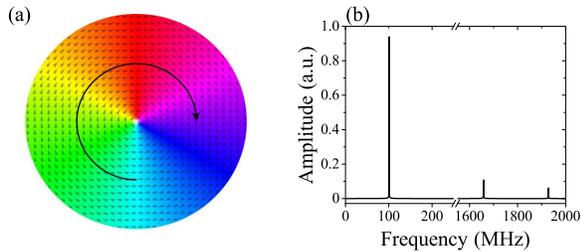}
\caption{(a) The spatial distribution of the vortex magnetization stabilized in a $r=180$ nm, $t=20$ nm nanodisc with $P=+1$ and $C=+1$. (b) Numerically calculated frequencies of the vortex state excitations with in-plane perturbation by using the FFT.
}\label{fig:2}
\end{figure}

As a specific example, we simulate a magnetic nanodisc composed of yttrium iron garnet (YIG) with $t=20$ nm, $r=180$ nm, and the results are shown in Fig.~\ref{fig:2}. Micromagnetic simulations with MuMax3 are used to estimate the results \cite{1.4899186}.
The magnetic structure of a vortex state is depicted in Fig.~\ref{fig:2}(a) with polarity $P=+1$ and chirality $C=+1$. It should be noted that the configuration corresponding to the polarity and circulation is irrelevant to our discussion, which can be specified as an initial condition.
Fig.~\ref{fig:2}(b) shows the excitation spectrum of the vortex by applying a pulse perturbation. Specifically, a magnetic field of $10$ $\mu$T is applied to the nanodisc in the lateral direction, leading to the precession of the vortex core. By removing the magnetic field, the vortex core oscillates back to its previous equilibrium position. The spectrum of the excited states can be obtained by performing the fast Fourier transform (FFT) to the averaged magnetization component of the vortex \cite{1.1450816}.
As revealed in Fig.~\ref{fig:2}(b), the first mode corresponding to the peak at $f_G\approx100$ MHz is the gyrotropic mode described previously, which is consistent with the analytical results (see more in appendix \ref{app:b}). Other frequencies of high-order azimuthal modes at $1.66$ GHz and $1.93$ GHz can also be identified, which are unrelated to the discussion. In addition, the vortex structure will be deformed by applying a uniform perpendicular field, resulting in a linear dependence of the gyrotropic frequency until threshold fields, which correspond to the reversal of the core or the saturation field according to the polarity direction \cite{PhysRevLett.102.177602,1.3373833}.

In this setup, the coupling of the magnetic vortex gyration to the mechanical resonator is originated by a ferromagnet mounted to the tip
of the cantilever \cite{PhysRevLett.99.140403}. The rod-shaped magnet transduces the mechanical vibration $a(\tau)$ of the cantilever along the y-axis into an oscillating magnetic field $\mathbf{B}_{r}(\tau)=G_{v}a(\tau)\mathbf{e}_z$ at the center of the disc, with the unit vector $\mathbf{e}_z$ pointing to the z-axis.
Approximating the nanorod by a magnetic dipole $\boldsymbol{\mu}_{m}$ oriented along the z-axis owing to the shape anisotropy, we get $G_{v}=3\mu_{0}\vert\boldsymbol{\mu}_{m}\vert/4\pi d_{vc}^4$, where $\mu_{0}$ is the vacuum permeability, and we can see that $G_{v}$ can be adjusted by changing the distance $d_{vc}$ with the given magnet.
The nanomechanical cantilever is described by $\hat{H}_{c}=\hbar \omega_{c}\hat{a}^{\dagger}_{c}\hat{a}_{c}$ (see more details in appendix \ref{app:a}) and the quantization of the magnetic field yields $\hat{B}_{r}=B_{vc}(\hat{a}_{c}+\hat{a}^{\dagger}_{c})$, where $B_{vc}=G_{v}a_{0}$, $\omega_{c}$ is the frequency of the cantilever, $\hat{a}_{c}$ ($\hat{a}^{\dagger}_{c}$) is the annihilation (creation) operator of the resonator mode, and $a_{0}=\sqrt{\hbar/2M\omega_{c}}$ is the amplitude of the zero-point fluctuation of the cantilever with mass $M$.
The nanomechanical resonator contains a complicated spectrum of vibrational modes, and the well-resolved fundamental flexural eigenmode can couple to the vortex gyration by utilizing a high-quality cantilever at frequency $\omega_{c}/2\pi\approx 0.56\sqrt{E/12\rho(1+c)}(t_{c}/l_{c}^2)$. Here, we take $t_{c}\leq w_{c}\ll l_{c}$ as the cantilever dimensions of thickness, width, and length, $E$ is Young's modulus, $\rho$ is the mass density, and the extra mass $m$ of the magnet with the supporting paddle at the tip is accounted for by $c=m/0.24\rho l_{c}w_{c}t_{c}$. Besides, by modeling the cantilever tip as a harmonic oscillator with an effective mass $M\approx 0.24\rho l_{c}w_{c}t_{c}+m$ at frequency $\omega_{c}$, we can compensate for the extra shift of the fluctuation amplitude introduced by the magnet. To match the gyrotropic mode frequency of the vortex, high frequencies of the cantilevers are necessary, which will lead to a major drawback of decreasing the quality factor $Q$. Consequently, the frequency of about a few hundred megahertz or lower is an acceptable range with present experimental conditions, which can achieve resonance with the gyrotropic mode of the vortex conveniently and has a relatively high $Q$.

Then we discuss the coupling between vortex excitations and the mechanical resonator. Specifically, the core of the vortex is positioned below the magnet and is subject to a transverse magnetic field, which can excite the gyrotropic mode. Considering the case that the resonator contains few or only one phonon, the dynamics of the vortex can be modeled as a harmonic-oscillator-like equation of motion due to the assumption that the magnetic texture is disrupted within the linear response regime, and can be described by the Hamiltonian $\hat{H}_{v}=\hbar \omega_{v}\hat{a}_{v}^{\dagger} \hat{a}_{v}$ with $\hat{a}_{v}$ ($\hat{a}_{v}^{\dagger}$) being a bosonic operator that annihilates (creates) the vortex excitations of the gyrotropic mode at frequency $\omega_{v}$ (see more details in appendix \ref{app:b}).
For a given material, the frequencies of vortices depend on the geometrical dimensions and can be slightly tuned by an out-of-plane field, thus enabling the resonance with a nanomechanical resonator $\omega_{v}=\omega_{c}$.
By calculating the Zeeman coupling between the magnetic moments and the magnetic field \cite{acsphotonics.8b00954,ab52d7}, the Hamiltonian of the quantum vortex gyration-phonon model is given as (see more details in appendix \ref{app:c})
\begin{equation}\label{equ:1}
\hat{H}_{vc}/\hbar =\omega_{c}\hat{a}_{c}^{\dagger}\hat{a}_{c}+\omega_{v}\hat{a}_{v}^{\dagger}\hat{a}_{v}+g_{vc}(\hat{a}_{v}^{\dagger}+\hat{a}_{v})(\hat{a}_{c}^{\dagger}+\hat{a}_{c}),
\end{equation}
and the coupling strength $g_{vc}$ is
\begin{equation}\label{equ:2}
g_{vc}\equiv\frac{B_{vc}}{2}\sqrt{\frac{VX}{\hbar}},
\end{equation}
where $V=\pi r^2t$ is the disc volume, $X=\xi^{2}M_{s}\gamma_{g}/2\pi$ with $M_{s}$ the saturation magnetization, $\gamma_{g}/2\pi=28$ GHz/T the gyromagnetic ratio and a geometrical factor $\xi=2/3$ in the case of discs. The coupling strength $g_{vc}$ is proportional to the magnetic field generated by the magnet as a result of the Zeeman type coupling \cite{acsphotonics.8b00954,ab52d7}. Similar to the case of the cavity-magnon coupling $g_{c-m}$, which is proportional to the spin number of the magnetic spheres, the coupling strength in our case shows a dependence on volume, which is related to the number of spins as well.

Taking into account the dissipations of the magnetic vortex $\gamma$ (see more details in appendix \ref{app:b}) and nanomechanical resonator $\kappa=\omega_{c}/Q$ with $Q$ as the quality factor, we can use the master equation to describe the full dynamics of the system, which has the form
\begin{equation}\begin{split}
\dot{\hat{\rho}}=&-\frac{i}{\hbar}[\hat{H}_{vc},\hat{\rho}]+(\bar{n}_{v}+1)\gamma\mathcal{D}[\hat{a}_{v}]\hat{\rho}+\bar{n}_{v}\gamma\mathcal{D}[\hat{a}_{v}^{\dagger}]\hat{\rho}\\
&+(\bar{n}_{c}+1)\kappa\mathcal{D}[\hat{a}_{c}]\hat{\rho}+\bar{n}_{c}\kappa\mathcal{D}[\hat{a}_{c}^{\dagger}]\hat{\rho},
\end{split}\end{equation}
with $\mathcal{D}[\hat{o}]\hat{\rho}=\hat{o}\hat{\rho}\hat{o}^{\dagger}-\{\hat{o}^{\dagger}\hat{o},\hat{\rho}\}/2$ for a specified operator $\hat{o}$ and $\bar{n}_{j}=[\exp(\hbar\omega_{j}/k_{B}T)-1]^{-1} (j=v, c)$ being the thermal occupation number at the environment temperature $T$, which is assumed as $T\approx 10$ mK in a dilution refrigerator. When the coupling strength exceeds both the vortex and resonator damping rates, the strong coupling regime emerges.

\section{Strong coupling between the phonon and the vortex excitation}\label{sec:3}
\begin{figure}[htbp]
\centering\includegraphics[width=0.45\textwidth]{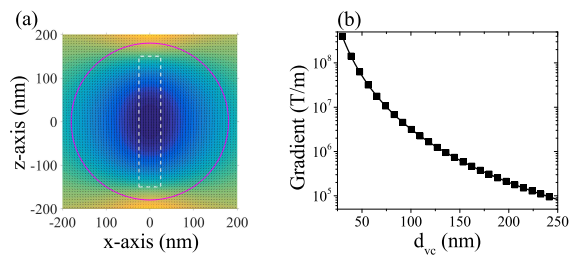}
\caption{(a) The magnetic field distribution of the magnet at $d=150$ nm. The purple circle and the dashed rectangle represent the projection position of the nanodisc and the magnet, respectively. (b) The magnetic field gradient as a function of distance. The dimensions of magnet is chosen as $(l_{m}, w_{m}, t_{m})=(0.3, 0.05, 0.05)$ $\mu$m, and the nanodisc radius $r=180$ nm.
}\label{fig:3}
\end{figure}

To realize coherent quantum dynamics, a strong coupling regime is necessary.
When considering the coupling $g_{vc}$, some additional situations should be taken into account. In practice, since the magnetic field is a crucial component of coupling, its distribution in space is inhomogeneous. Specific to our setup, the disc response to the low-amplitude driving field, i.e., the shift from the equilibrium position, is at a small range around the disc center, which allows us to consider the dependence only on the magnetic field at the center of discs.
An example of the field distribution in the x-z plane \cite{1.1883308} is shown in Fig.~\ref{fig:3}(a). The white dashed rectangle represents a magnet, and the purple circle represents the disc. The magnetic field is evenly distributed in the center region of the disc, as shown in the figure, making our assumption a good approximation.
The amplitude of the magnetic field at the center $\mathbf{r}$ has the form $B_{vc}=G_{v}a_{0}$, in which $G_{v}$ varies with the distance between the magnet and the vortex core. Fig.~\ref{fig:3}(b) shows the dependence of gradient $G_{v}$ on distance $d_{vc}$, and the gradient drops sharply as the distance increases.

The material and geometrical dimensions of the nanodisc are also critical for the coupling strength, and the selection of these parameters is related to the resonance condition with a mechanical resonator. To reach the strong coupling regime, it is evident to select a material with low damping characteristics. Furthermore, unlike the vortex and coplanar waveguide cavity coupling, the frequency of the mechanical resonator is relatively low, therefore, it is preferable to have lower saturation magnetization materials, which show less geometrical confinement.
For specific instance, YIG would be an appropriate choice, since it exhibits one of the lowest Gilbert damping parameters with $\alpha_{LLG}\approx5\times10^{-5}$, and low saturation magnetization $\mu_{0}M_{s}\approx 0.18$ T \cite{lmag.2020.2989687,pssb.201900644,nphys3770,acsphotonics.8b00954}, leading to an eigenfrequency of gyrotropic mode in the range of about tens to hundreds of megahertz.
Another suitable material is CoFe, which also has a low damping parameter $\alpha_{LLG}\approx5\times10^{-4}$ \cite{nphys3770}, but with large $\mu_{0}M_{s}\approx 2.4$ T, the aspect ratio $\beta$ should be small enough, which might necessitate a large radius of the disc or the utilization of a resonator with a high frequency in sub-gigahertz.

\begin{figure}[htbp]
\centering\includegraphics[width=0.45\textwidth]{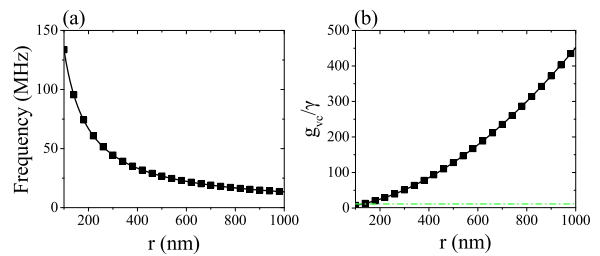}
\caption{(a) The eigenfrequencies of the gyrotropic mode as a function of the radius $r$ of the disc, which is made by YIG. (b) Coupling strength $g_{vc}$ normalized to the damping rate of the vortex versus the disc radius. The green dashed-dotted line indicates the value of $1$.
}\label{fig:4}
\end{figure}

Here we take YIG as the material and investigate the performance under various conditions. For convenience, we assume $t=15$ nm, $G_{v}=5\times10^{5}$ T/m \cite{nnano.2007.105,PhysRevB.79.041302,PhysRevLett.125.153602} and $a_{0}=0.5\times10^{-13}$ m, which is a little bit larger than the experiment \cite{j.physrep.2011.12.004,nnano.2006.208}. This magnetic gradient is equivalent to the amplitude at a distance of $160$ nm with the magnet parameters given in Fig.~\ref{fig:3}.
The variations of the gyrotropic frequency of the YIG disc with the radius at a fixed thickness are shown in Fig.~\ref{fig:4}(a) as well as the ratio $g_{vc}/\gamma$ under corresponding dimension conditions in Fig.~\ref{fig:4}(b). It reveals that increasing the radius of the nanodisc allows for the realization of frequencies close to the frequency range of  nanomechanical resonators, as well as an increase in the ratio of the coupling strength to the vortex damping.

\begin{figure}[htbp]
\centering\includegraphics[width=0.45\textwidth]{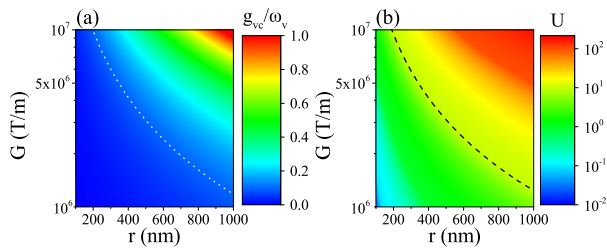}
\caption{The contour map of the coupling strength $g_{vc}$ normalized to the resonance frequency $\omega_{v}$ (a) and the USC measure $U$ (b) versus the disc radius and the gradient $G$. In Fig. (a) the dotted curves represent the value of $0.1$, while the dashed curve in Fig. (b) is $10$. The area with larger values than the curves indicates the USC condition. The other parameters are $t=15$ nm using YIG as the material and the quality factor of the cantilever $Q=1000$ with $a_{0}=0.5\times10^{-13}$ m.
}\label{fig:5}
\end{figure}

Considering further that the vortex approaches the magnet, i.e., the magnetic field gradient $G_{v}$ increases, the coupling strength of the system may achieve the USC regime. Fig.~\ref{fig:5} presents the dependence of the coupling strength $g_{vc}$ normalized to the vortex gyration frequency $\omega_{v}$ and parameter $U$ as a function of the radius $r$ and magnetic field gradient $G$ at the center of the vortex. The parameter $U=\sqrt{Cg_{vc}/\omega_{v}}$ is a measure of coherence \cite{RevModPhys.91.025005} that corresponds to the geometric mean value between the cooperativity $C=g_{vc}^{2}/\gamma\kappa$ and the ratio of the coupling strength to the resonance frequency. It is feasible to reach the unconventional properties of the USC regime under the condition $U\gg 1$, while $g/\omega\simeq0.1$ is usually marked as the beginning of the USC regime. The result in Fig.~\ref{fig:5}(a) shows that when the gradient $G$ is larger than about $10^7$ T/m, the USC regime can be easily achieved. Taking the dissipation into consideration, a similar result can be obtained from Fig.~\ref{fig:5}(b), where $U$ can reach a considerable value with the optimized parameters, far greater than unity.
Since the response of the magnetic disc is only affected by the magnetic field at the center, a larger disc radius will increase the number of spins and hence the coupling strength without causing side effects. Moreover, the coupling strength is proportional to the gradient, which is generated by the magnet. Though the gradient weakens rapidly with the distance, the magnitude within a particular distance is rather significant, as shown in Fig.~\ref{fig:3}(b). Consequently, we can achieve the requisite coupling regime by choosing the proper parameters.

\section{Application}\label{sec:4}
\begin{figure}[htbp]
\centering\includegraphics[width=0.45\textwidth]{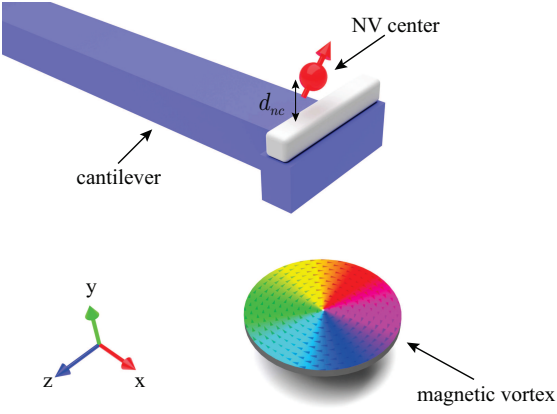}
\caption{Schematic of the tripartite hybrid system, where an extra NV center is positioned above the magnet of the proposed system at a distance of $d_{nc}=40$ nm. The parameters of the vortex-cantilever system are the same as in Fig.~\ref{fig:1}.
}\label{fig:6}
\end{figure}

The interaction of Hamiltonian~(\ref{equ:1}), when achieving the strong coupling regime, allows the coherent quantum state transfer between the gyrotropic mode of the vortex and the resonator mode. In this section, we discuss a realization of the quantum state transfer between a spin and the vortex excitation mode, where the vibration mode of the nanomechanical resonator serves as a quantum data bus. Here we arrange an NV center and a disc on either side of the magnet at the end of the cantilever, as shown in Fig.~\ref{fig:6}, where the direct interaction between the spin and the vortex is negligible due to the far distance \cite{nnano.2016.63,ncomms11584}.

The spin-vortex coupling is triggered by the magnetic field which is mainly formed by the vertical magnetization of the vortex core. To get a rough estimate of the coupling strength, we assume that the magnetization direction of the vortex is along the $y$-axis uniformly for simplification, which would lead to an overestimate of the coupling. Then we can approximate the field $\mathbf{B}$ from the vortex as a dipole field centered in the vortex core, and quantize the vortex as the Kittel mode which has the form $\hat{M}=m_{v}(\hat{a}_{v}+\hat{a}_{v}^{\dagger})$, with $m_{v}=\sqrt{\hbar\gamma_{g} M_{s}/(2V)}$ being the zero-point magnetization \cite{PhysRevB.105.075410,PhysRevB.101.125404,PhysRevA.103.043706}. The quantized magnetic field $\hat{B}$ right above the vortex core may then be obtained as
\begin{equation}
\hat{\mathbf{B}}=\frac{\mu_{0}m_{v}V}{2\pi y^{3}}(\hat{a}_{v}+\hat{a}_{v}^{\dagger})\mathbf{e}_{y},
\end{equation}
with $y$ being the distance between the NV center and the vortex core. The vortex-spin interaction in the dressed state basis of the spin can be described as
\begin{equation}
\hat{H}_{vn}=\hbar g_{vn}(\hat{a}_{v}+\hat{a}_{v}^{\dagger})(\hat{S}_{+}+\hat{S}_{-}),
\end{equation}
where $\hbar g_{vn}=\mu m_{v}V/y^3$ is the coupling strength, with $\mu=\mu_{0}\mu_{B}g_{s}/2\pi$, $g_{s}\simeq 2$ the land\'{e} factor, $\mu_{B}$ the Bohr magneton, and the spin operators $\hat{S}_{\pm}$ of the dressed states, which are obtained in the same way as the discussion below.
Due to the long-range distortion of the vortex domain, the field diminishes as about $y^{-3}$ with tiny fields lingering beyond the core at tens of nanometers, resulting in rapid attenuation of coupling strength with distance. Taking previous vortex parameters as an example, the coupling strength at about $y=200$ nm in our setup is $g_{vn}/2\pi\approx15$ kHz, which is much lower than the coupling strength of the vortex-phonon coupling and spin-phonon coupling (about several hundred megahertz), as well as the damping rate of the vortex, although this result gives the upper limit of the coupling strength. Furthermore, if only the vortex core is considered while the other part is magnetized in the in-plane direction, the coupling strength would be further reduced by about an order of magnitude. Accordingly, this interaction between the vortex and the NV center can be neglected in our setup.

The NV center \cite{j.physrep.2013.02.001} is a common quantum spin system consisting of a substitutional nitrogen-lattice vacancy pair replacing carbon atoms. The electronic ground state of a single NV center is a spin-triplet ground state, with a zero-splitting $D=2\pi\times2.87$ GHz between the degenerate sublevels $\vert m_{s}=\pm1\rangle$ and $\vert m_{s}=0\rangle$.
The crystalline axis of the NV center is referred to as the z-axis for convenience.
Typically, we apply a magnetic field $\mathbf{B}_{z}=B_{z}\mathbf{e}_{z}$ to the NV center to remove the degeneracy of the states $\vert m_{s}=\pm1\rangle$, while using $\mathbf{B}_{dr}=B_{0}\cos\omega_{0}t\mathbf{e}_{x}$ to drive the Rabi oscillations between $\vert m_{s}=0\rangle$ and the excited states $\vert m_{s}=\pm1\rangle$.
Considering a spin-mechanical setup \cite{PhysRevB.79.041302,Nphys2070}, a single NV center can magnetically couple to the mechanical motion of the cantilever through the
magnetic field gradient generated by the magnet on the tip and the Hamiltonian can be described in the dressed state basis.
Specifically (see more details in appendix \ref{app:c}), by defining the bright state $\vert B\rangle=(\vert +1\rangle+\vert -1\rangle)/\sqrt{2}$ and dark state $\vert D\rangle=(\vert +1\rangle-\vert -1\rangle)/\sqrt{2}$, we find that the state $\vert 0\rangle$ couples to the bright state $\vert B\rangle$ while the dark state $\vert D\rangle$ is decoupled. Then by switching to the dressed state basis \{$\vert G\rangle=\cos\theta\vert 0\rangle-\sin\theta\vert B\rangle$, $\vert E\rangle=\sin\theta\vert 0\rangle+\cos\theta\vert B\rangle$\}, the Hamiltonian of the system can be simplified under the condition $\Delta\gg\Omega$ with $\Delta$ and $\Omega$ being the detuning and Rabi frequency of the NV center sublevels, respectively. As a result, $\hat{H}_{nc}$ can be described by the JC interaction under the RWA, which has the form
\begin{equation}
\hat{H}_{nc}/\hbar=\omega_{c}\hat{a}_{c}^{\dagger}\hat{a}_{c}+\frac{1}{2}\Lambda\hat{\sigma}_{z}+g_{nc}(\hat{\sigma}_{+}\hat{a}_{c}+\hat{\sigma}_{-}\hat{a}_{c}^{\dagger}),
\end{equation}
where $\hbar g_{nc}=g_{s}\mu_{B}G_{nc}a_{0}$ is the coupling strength between an NV center and the cantilever with $G_{nc}=3\mu_{0}\vert\boldsymbol{\mu}_{m}\vert/4\pi d_{nc}^4$ the magnetic field gradient at the position of the NV center, $d_{nc}$ is their distance, $\hat{\sigma}_{-}=\vert D\rangle\langle E\vert$, $\hat{\sigma}_{+}=\hat{\sigma}_{-}^{\dagger}$, $\hat{\sigma}_{z}=\vert E\rangle\langle E\vert-\vert D\rangle\langle D\vert$, $\tan2\theta=2\sqrt{2}\Omega/\Delta$, $\Lambda=\frac{2\Omega^{2}}{\Delta}$.

In the rotating frame at the NV center frequency $\Lambda$ and under the RWA, the whole Hamiltonian including the vortex gyration-cantilever coupling reads
\begin{eqnarray}
 \hat{H}/\hbar&=&\Delta_{1}\hat{a}_{c}^{\dagger}\hat{a}_{c}+\Delta_{2}\hat{a}_{v}^{\dagger}\hat{a}_{v}+g_{vc}(\hat{a}_{v}^{\dagger}\hat{a}_{c}+\hat{a}_{v}\hat{a}_{c}^{\dagger})\nonumber\\
 &&+g_{nc}(\hat{\sigma}_{+}\hat{a}_{c}+\hat{\sigma}_{-}\hat{a}_{c}^{\dagger}),
\end{eqnarray}
with $\Delta_{1}\equiv\omega_{c}-\Lambda$ and $\Delta_{2}\equiv\omega_{v}-\Lambda$.
By eliminating the mechanical mode $\hat{a}_{c}$ for large detuning condition $\vert\Delta_{1}\vert\gg g_{vc}, g_{nc}$, we can acquire an effective vortex gyration-spin interaction while there is no or weak direct interaction between them \cite{PhysRevLett.112.213602}. The effective Hamiltonian can be written as
\begin{equation}\begin{split}\label{equ:6}
\hat{H}_{eff}/\hbar=&(\Delta_{2}-\beta^{2}\Delta_{1})\hat{a}_{v}^{\dagger}\hat{a}_{v}-\frac{1}{2}\alpha^{2}\Delta_{1}\hat{\sigma}_{z}\\
&+g_{eff}(\hat{a}_{v}\hat{\sigma}_{+}+\hat{a}_{v}^{\dagger}\hat{\sigma}_{-}),
\end{split}\end{equation}
where $\alpha=g_{nc}/\vert\Delta_{1}\vert$ and $\beta=g_{vc}/\vert\Delta_{1}\vert$ represent the dimensionless interaction parameters. Other parameters are described as follow: $g_{eff}=\beta g_{nc}$, $\kappa_{eff}=\kappa_{2}+\beta^{2}\kappa_{1}$, and $\gamma_{eff}=\gamma+\alpha^{2}\kappa_{1}$; here we denote the dissipation of the cantilever and NV center as $\kappa_{1}$ and $\kappa_{2}$, respectively. It can be seen that for $\alpha, \beta\ll 1$ in the effective Hamiltonian, the dissipation caused by the mechanical resonator is greatly reduced, which is helpful to achieve the strong coupling regime.

\begin{figure}[htbp]
\centering\includegraphics[width=0.45\textwidth]{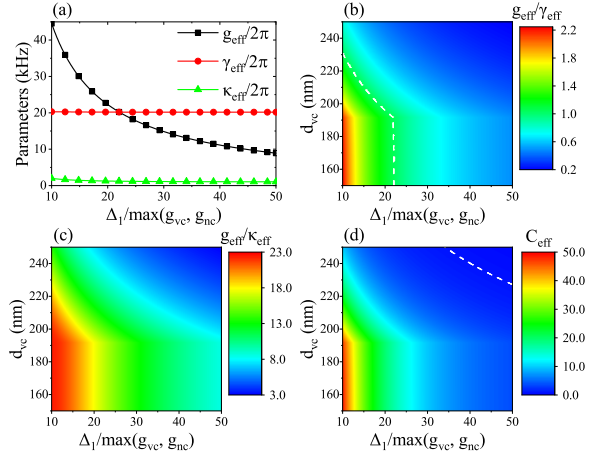}
\caption{(a) Parameters $g_{eff}/2\pi$, $\gamma_{eff}/2\pi$ and $\kappa_{eff}/2\pi$ as a function of detunings. (b)-(d) Contour plots of the ratio $g_{eff}/\gamma_{eff}$, $g_{eff}/\kappa_{eff}$ and the cooperativity $C_{eff}$ as a function of the detuning and distance $d_{vc}$. Due to the requirement of the large detuning condition, $\Delta_{1}$ should compare with the larger coupling strength, i.e., $\max(g_{vc},g_{nc})$, since $g_{vc}$ become smaller than $g_{nc}$ at $d_{vc}\approx190$ nm with this configuration. This change can also be revealed by the results in (b)-(d), where the boundaries change from straight lines to curves. The white dashed lines indicate the value of 1, which corresponds to the strong coupling.
}\label{fig:7}
\end{figure}
In Fig.~\ref{fig:7}, we calculate the effective parameters $g_{eff}/2\pi$, $\gamma_{eff}/2\pi$ and $\kappa_{eff}/2\pi$  as a function of the detuning $\Delta_{1}$. The parameters are $r=180$ nm, $t=20$ nm, $d_{vc}=150$ nm, $d_{nc}=40$ nm, and the cantilever dimension $(l_{c}, w_{c}, t_{c})=(1.2, 0.2, 0.15)$ $\mu$m with a magnet $(l_{m}, w_{m}, t_{m})=(0.3, 0.05, 0.05)$ $\mu$m. With these parameters, we can obtain the coupling strengths of vortex-phonon and spin-phonon interaction, which are $g_{vc}/2\pi\simeq 1.2$ MHz and $g_{nc}/2\pi\simeq 0.45$ MHz, respectively, and the resonance frequency $\omega_{v}/2\pi=\omega_{c}/2\pi\simeq 100$ MHz. For the dissipation, the vortex damping rate is $\gamma/2\pi\simeq 20$ kHz, and the quality factor of the cantilever is $Q=1000$, resulting in the damping rate $\kappa_{1}/2\pi\simeq 100$ kHz, and the dephasing rate of NV centers is $\kappa_{2}/2\pi=1$ kHz. Fig.~\ref{fig:7}(a) reveals that with a suitable detuning $\Delta_{1}$, the effective coupling strength $g_{eff}$ can exceeds both the effective damping rate $\gamma_{eff}$ and $\kappa_{eff}$, we can also get a coupling strength $g_{eff}/2\pi\approx 40$ kHz with $\Delta_{1}=10g_{nc}$, which is greater than the upper limit of the direct coupling strength. This result is valid for the large damping rate of the cantilever as well.
For a deeper understanding of the parameter ranges, in Figs.~\ref{fig:7}(b)-(d), we also plot the ratio $g_{eff}/\gamma_{eff}$, $g_{eff}/\kappa_{eff}$ and the cooperativity parameter $C_{eff}\equiv g_{eff}^{2}/\gamma_{eff}\kappa_{eff}$ as a function of $\Delta_{1}$ and $d_{vc}$. In order to meet the large detuning condition, the detuning $\Delta_{1}$ should be greater than both of the coupling strengths, namely, the detuning needs to compare with the larger coupling since $g_{vc}$ changes with distance $d_{vc}$.
It shows that the ratios and the cooperativity parameters increase as the coupling strength $g_{vc}$ gets larger (manifested as a reduction in the distance $d_{vc}$) till a fixed value when $g_{vc}=g_{nc}$, and decrease when the detuning gets larger. A value larger than $1$ indicates that the system has achieved the strong coupling regime.

\begin{figure}[htbp]
\centering\includegraphics[width=0.45\textwidth]{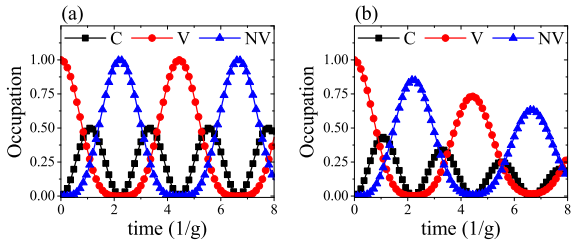}
\caption{Time evolution of each part occupations in the hybrid system, without (a) and with (b) dissipations. Here $g_{vc}=g_{nc}=0.45\times2\pi$ MHz and other parameters are the same as those in Fig. \ref{fig:7}, i.e. $\gamma\simeq 0.045g$, $\kappa_{1}\simeq 0.222g$, $\kappa_{2}\simeq 0.002g$. We use C, V, and NV to represent the occupations of the cantilever, the vortex gyration, and the NV center, respectively, and V is initialized in the excited state while C and the NV are in the vacuum and ground state.
}\label{fig:8}
\end{figure}

In order to verify Hamiltonian~(\ref{equ:6}), we numerically simulate the system to obtain the occupations in the time domain with $g=g_{vc}=g_{nc}=0.45\times2\pi$ MHz and assume that the vortex gyration is in the excited state while the cantilever and the NV center are in the vacuum and ground state, respectively.
Firstly, the simulation results of the original Hamiltonian $\hat{H}$ are shown in Fig.~\ref{fig:8}, in which the Rabi oscillation phenomena on the microsecond scale ($1/g$) can be observed in Fig.~\ref{fig:8}(a) without loss.
As the dynamics proceeds, the quantum state transfers from the vortex gyration to the NV center through the cantilever with the same amplitude and then vice versa. The cantilever here serves as a quantum bus for the transition of the state, as we can see from the transfer route. The occupations of the cantilever, the vortex gyration, and the NV center are represented by C, V, and NV, respectively. When considering the effect of dissipations, the amplitudes of the occupations will decay with the dynamic evolution, as shown in Fig.~\ref{fig:8}(b). The dissipation parameters are the same as those in Fig.~\ref{fig:7}.

\begin{figure}[htbp]
\centering\includegraphics[width=0.45\textwidth]{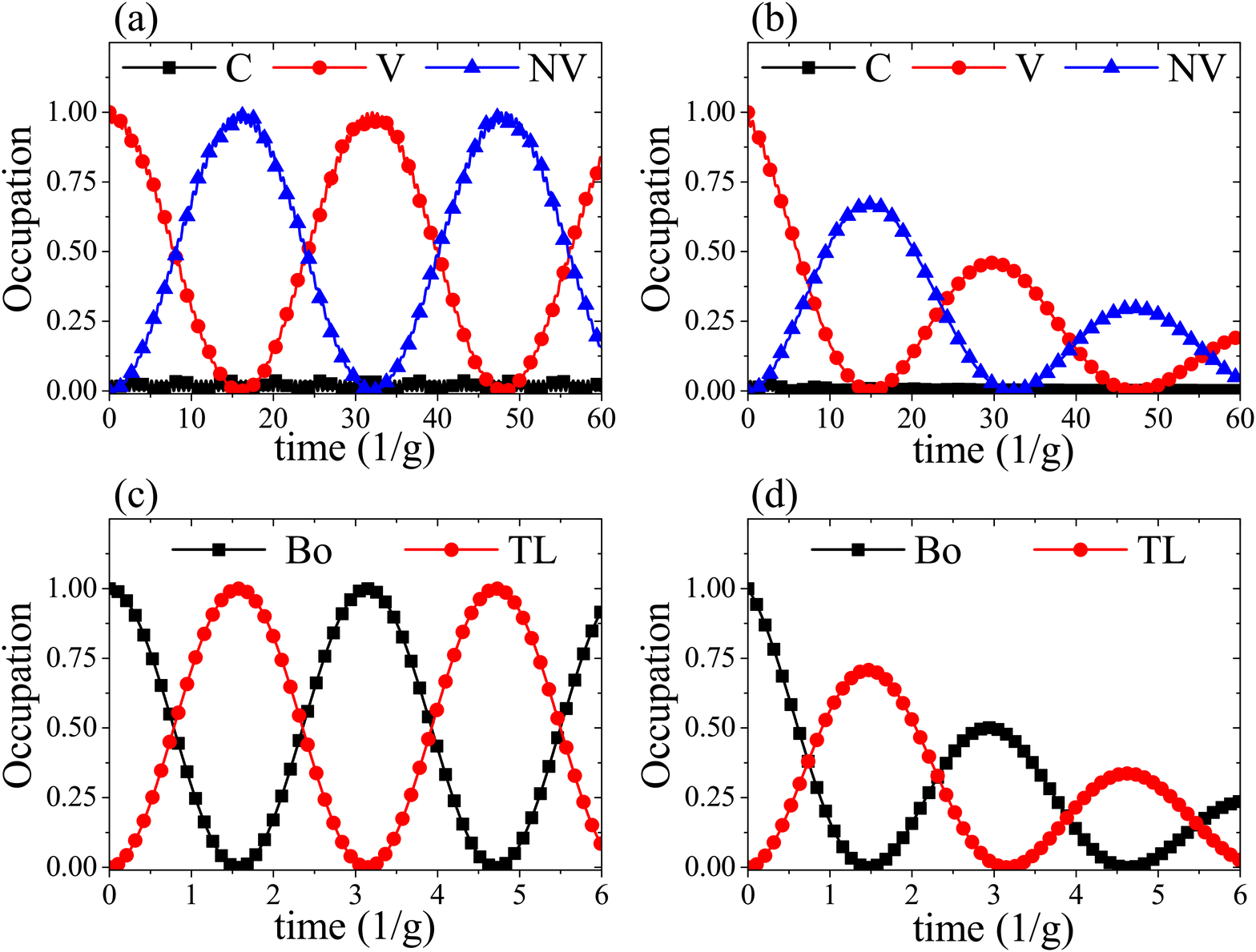}
\caption{(a) and (b) Time evolution of the occupations under the large detuning condition in this system. The parameters are the same as in Fig.~\ref{fig:9}, i.e. $\gamma\simeq 0.045g$, $\kappa_{1}\simeq 0.222g$, $\kappa_{2}\simeq 0.002g$. (c) and (d) Contradistinction of the occupation evolution of a bipartite system over time, and the dissipation of the bosonic mode and the two-level system are chosen as $K_{1}\simeq 0.45G$, $K_{2}\simeq 0.02G$ with the coupling strength $G/2\pi=0.45$ MHz. In (a) and (c) the dissipations are neglected while (b) and (d) are considered. Here the bosonic mode and the two-level system are represented as Bo and TL, respectively.
}\label{fig:9}
\end{figure}
The dynamics under the large detuning condition ($\Delta_{1}\gg g$) are then considered. Figs.~\ref{fig:9}(a) and (b) demonstrate the occupations with and without dissipations, respectively. Unlike the previous condition, the Rabi oscillations of the vortex gyration and NV center occupation are plainly evident, while the cantilever excitation can be ignored. This result indicates an effective interaction between the vortex gyration and the NV center brought on by the virtual excitation of the cantilever, and the coherent time is sufficient for the energy exchange, as shown in Fig.~\ref{fig:9}(b). As a comparison, we depict the JC model occupations of a direct coupling setup between a bosonic mode (marked as Bo) and a two-level system (marked as TL) in Figs.~\ref{fig:9}(c) and (d), which demonstrate that the effective Hamiltonian is a decent approximation.

\section{Conclusions}\label{sec:6}
In summary, we propose a hybrid system where the gyrotropic mode of a magnetic vortex can couple to the phonon in a nanomechanical resonator through a magnetic field gradient. By adjusting the disc geometries and the distance between the vortex and the resonator, the coupling strength can be increased from the strong coupling to the USC regime. The magnetic vortex in this system is extremely stable due to the topological protection, which can be adapted to other texture excitations, such as skyrmions, and cantilevers with a sharp magnetic tip or doubly clamped beams with a magnet are also valid. We also provide an application that enables the indirect coupling between a magnetic vortex and an NV center through virtual phonon excitations of the cantilever under the large detuning condition. This hybrid quantum system may facilitate the investigations of the USC regime and macroscopic quantum physics, as well as provide a novel platform for quantum information processing.

\begin{acknowledgments}
This work was supported by the National Natural Science Foundation of China under Grant No. 92065105, and the Natural Science Basic Research Program of Shaanxi (Program No. 2020JC-02). The simulations of the occupation part are coded in PYTHON using the QuTiP library \cite{j.cpc.2012.02.021,j.cpc.2012.11.019}.
\end{acknowledgments}

\appendix
\section{Quantization of the nanomechanical resonator}\label{app:a}
The tip of the nanomechanical cantilever can be treated as a simple harmonic oscillator when it vibrates. We can describe it with the Hamiltonian $\hat{H}_{c}$, which is formed from the sum of the kinetic and potential energies and has the form
\begin{equation}
\hat{H}_c=\frac{\hat{p}^2}{2M}+\frac{M\omega_{c}^2}{2}\hat{r}^2,
\end{equation}
with $M$ being the mass of the cantilever, $\omega_{c}$ being the eigen frequency of the cantilever, $\hat{p}$, and $\hat{r}$ being the position and momentum operators which have the commutation relation $[\hat{r},\hat{p}]=i\hbar$. Here the mass $M\approx 0.24\rho l_{c}w_{c}t_{c}+m$ is a result of the integration of the mode function. The eigenstates of the Hamiltonian are $\vert n\rangle$, with the relation $\hat{H}_c\vert n\rangle=(n+\frac{1}{2})\hbar\omega_{c}\vert n\rangle$, where $n$ is an integer.
Using the commutation relations of the Hamiltonian $\hat{H}_c$ with the position and momentum operator, we can build the raising (creation) operator $\hat{a}_{c}^{\dagger}$ and the lowering (annihilation) operator $\hat{a}_{c}$:
\begin{equation}\begin{split}
\hat{a}_c^{\dagger}&=\sqrt{\frac{M\omega_{c}}{2\hbar}}\hat{r}-i\frac{1}{\sqrt{2\hbar M\omega_{c}}}\hat{p},\\
\hat{a}_c&=\sqrt{\frac{M\omega_{c}}{2\hbar}}\hat{r}+i\frac{1}{\sqrt{2\hbar M\omega_{c}}}\hat{p}.
\end{split}\end{equation}
These operators have the following effects on the eigenstates $\vert n\rangle$ of $\hat{H}_{c}$
\begin{equation}\begin{split}
\hat{a}_c^{\dagger}\vert n\rangle&=\sqrt{n+1}\vert n+1\rangle,\\
\hat{a}_c\vert n\rangle&=\sqrt{n}\vert n-1\rangle,
\end{split}\end{equation}
and furthermore, the Hamiltonian $\hat{H}_c$ can be written in the standard form of harmonic oscillator
\begin{equation}
\hat{H}_{c}=\hbar\omega_{c}\hat{a}_{c}^{\dagger}\hat{a}_{c}.
\end{equation}

\section{The frequency of the vortex gyrotropic mode and numerical simulations}\label{app:b}
By solving the Thiele equation, the gyrotropic mode frequency of a vortex can be obtained, which is mainly determined by magnetostatics. The gyrotropic eigenfrequency $\omega_{v}$ at zero field is directly proportional to the aspect ratio $\beta=t/r$ for small values \cite{PhysRevLett.96.067205,jnn.2008.003}, which has the form
\begin{equation}
\omega_{v}=\frac{10}{9}\frac{\gamma_{g}\mu_{0}M_{s}\beta}{2\pi},
\end{equation}
with $\gamma_{g}$ the gyromagnetic ratio, $\mu_{0}$ the vacuum permeability and $M_{s}$ the material’s saturation magnetization. We can also get the line width with the equation
\begin{equation}
\gamma=4\pi\alpha_{LLG}[1+\ln(r/r_{v})/2],
\end{equation}
where $\alpha_{LLG}$ is the Gilbert damping parameter, $r_{v}\sim 1.58\lambda_{L}(t/\lambda_{L})^{1/3}$ is radius of vortex core and
$\lambda_{L}$ is the exchange length which has the form $\lambda_{L}=\sqrt{2A/\mu_{0}M_{s}^{2}}$ with $A$ the exchange stiffness.
Taking YIG as an example, the parameters are as follows: $\alpha_{LLG}\approx5\times 10^{-5}$, $\mu_{0}M_{s}\approx0.18$ T and $A=1.9$ pJ/m.
CoFe is another potential material in this scheme, which also has a low Gilbert damping parameter. But with a much larger saturation magnetization, the frequency of gyrotropic mode is higher, which might limit the coupling with a nanomechanical resonator. The specific parameters are $\alpha_{FeCo}\approx5\times 10^{-4}$, $\mu_{0}M_{FeCo}\approx2.4$ T and $A_{FeCo}=26$ pJ/m.

A micromagnetic simulation program Mumax3 is used to execute the numerical simulations. This software employs a finite-difference discretization to solve the time and space dependent magnetization evolution of nanoscale microscale magnets. The required material parameters include the saturation magnetization, Gilbert damping parameter, exchange stiffness constant, which are chosen as $\mu_{0}M_{s}=0.18$ T, $\alpha_{LLG}=5\times10^{-5}$, $A=1.9\times10^{-12}$ J/m, respectively when YIG is selected. For the geometric parameters, we choose a nanodisc with radius $r=180$ nm and thickness $t=20$ nm, which is placed in a box. The box is discretized into $128\times 128\times 8$ identical cells, and the unit cell element size is $3\times 3\times 2.8$ nm$^{3}$, resulting in a $384 \times 384\times 22.4$ nm$^3$ box. The texture of the magnetic vortex ground energy can be obtained after relaxing the YIG nanodisc.

The dynamics of the magnetic vortex are characterized using a perturbation field. Specifically, a magnetic field $B_{ext}=10$ mT is first applied to the remanent vortex distribution in the in-plane direction, leading to the vortex core precession. The driving field is then removed after $200$ ns, and the average magnetizations are recorded \cite{1.1450816}, which allows us to utilize FFT to determine the frequencies of the gyrotropic modes. The results are depicted in Fig.~\ref{fig:2}(b), where the first peak $f\approx 100$ MHz corresponds to the gyrotropic mode we need, and the value is in agreement with the theoretical result \cite{PhysRevLett.96.067205}.

\section{The interaction between a magnetic vortex and a nanomechanical resonator}\label{app:c}
In this appendix, we show the derivation of the coupling strength $g_{vc}$ in equation~(\ref{equ:2}). The interaction between a magnetic field and the spins in the disc is of Zeeman type, which has the form
\begin{equation}
\hat{H}_{vcI}=\sum_{i}\mu_{i}B_{r},
\end{equation}
where $\mu_{i}$ is the magnetic dipole of the $i$-spin, and we approximate the fluctuations of the magnetic field with the value at the center of the disc $B_{r}$, as discussed in the main text. Here we only consider the $z$-component. By using the method in appendix A, we can obtain the quantization of the magnetic field, while the quantized form of the vortex magnetization can be written by using the collective variable to describe the vortex precession. Then the Hamiltonian of the interaction term has the form
\begin{equation}
\hat{H}_{vcI}=VmB_{vc}(\hat{a}_{c}^{\dagger}+\hat{a}_{c})(\hat{a}_{v}^{\dagger}+\hat{a}_{v}),
\end{equation}
with $V=\pi r^2t$ the disc volume, $m$ the normalized magnetization, and $\hat{a}_{v}$ ($\hat{a}_{v}^{\dagger}$) being the creation (annihilation) operator of the vortex. Thus the coupling strength is
\begin{equation}\label{equ:C3}
\hbar g_{vc}=VmB_{vc}.
\end{equation}
This interaction indicates that the vortex is driven by the magnetic field with average $\hbar g_{vc}\langle \hat{a}_{c}^{\dagger}+\hat{a}_{c}\rangle$. Considering the single quantum limit, we can use the replacement $\langle\hat{a}_{c}^{\dagger}+\hat{a}_{c}\rangle=2\cos(\omega_{c}t)$, resulting in the maximum magnetization response of the vortex $\Delta M$ with the form
\begin{equation}
\Delta M=\frac{8\pi mg_{vc}}{\gamma}.
\end{equation}
Using the equation~\ref{equ:C3} and the vortex susceptibility expression $\chi=\Delta M/B_{vc}=\gamma_{g}\xi^{2}M_{s}/\gamma$, we arrive at the coupling strength
\begin{equation}
g_{vc}=\frac{B_{vc}}{2}\sqrt{\frac{V\xi^{2}M_{s}\gamma_{g}}{2\pi\hbar}}.
\end{equation}

\section{The interaction between an NV center and a nanomechanical resonator}\label{app:d}
Under the influence of the various external magnetic fields, the NV center can be described by the Hamiltonian
\begin{equation}
\hat{H}_{NV}=\hbar D\hat{S}_{z}^2+\mu_{B}g_{s}\mathbf{B}_{z}\hat{S}_{z}+\mu_{B}g_{s}(\mathbf{B}_{dr}+\mathbf{B}_{nc})\cdot\hat{\mathbf{S}},
\end{equation}
with $\mu_{B}$ the Bohr magneton, $g_{s}\simeq 2$ the land\'{e} factor and $\hat{\mathbf{S}}$ the spin operator of the NV center. The magnetic field $\mathbf{B}_{z}=B_{z}\mathbf{e}_{z}$ is applied to remove the degeneracy of the states $\vert\pm1\rangle$, while the driving field $\mathbf{B}_{dr}=B_{0}\cos\omega_{0}t\mathbf{e}_{x}$ is added to drive the Rabi oscillations. $\mathbf{B}_{nc}=G_{nc}a_{0}\mathbf{e}_{z}$ is the field generated by the oscillation of the magnet at the cantilever tip.
In the basis \{$\vert m_s\rangle, m_{s}=0, \pm1$\}, which is defined by the eigenstates of $\hat{S}_{z}$ with $\hat{S}_{z}\vert m_s\rangle=m_{s}\vert m_s\rangle$, we have
\begin{equation}\begin{split}
\hat{H}_{NV}=&\sum_{m_{s}}[\langle m_{s}\vert(\hbar D\hat{S}_{z}^2+\mu_{B}g_{s}B_{z}\hat{S}_{z})\vert m_{s}\rangle]\vert m_{s}\rangle\langle m_{s}\vert\\
&+\sum_{m_{s},m_{s}'}\mu_{B}g_{s}B_{0}\cos\omega_{0}t\langle m_{s}\vert\hat{S}_{x}\vert m_{s}'\rangle\vert m_{s}\rangle\langle m_{s}'\vert\\
&+\sum_{m_{s}}\mu_{B}g_{s}G_{nv}a_{0}\langle m_{s}\vert\hat{S}_{z}\vert m_{s}\rangle\vert m_{s}\rangle\langle m_{s}\vert(\hat{a}_{c}+\hat{a}_{c}^{\dagger}).
\end{split}\end{equation}
Taking $\mathbf{B}_{dr}=B_{0}(e^{i\omega_{0}t}+e^{-i\omega_{0}t})\mathbf{e}_{x}/2$ into account and in the rotating-frame at $\omega_{0}$, we can obtain $\hat{H}_{NV}$ under RWA
\begin{equation}\begin{split}
\hat{H}_{NV}\approx&\hbar\Delta_{+}\vert +1\rangle\langle +1\vert+\hbar\Delta_{-}\vert -1\rangle\langle -1\vert\\
&+\hbar\Omega(\vert +1\rangle\langle 0\vert+\vert -1\rangle\langle 0\vert+H.c.)\\
&+\hbar g_{nc}(\vert +1\rangle\langle +1\vert-\vert -1\rangle\langle -1\vert)(\hat{a}_{c}+\hat{a}_{c}^{\dagger}),
\end{split}\end{equation}
with $\hbar\Delta_{\pm}=\hbar D\pm\mu_{B}g_{s}B_{z}-\hbar\omega_{0}$, $\hbar\Omega=\frac{\sqrt{2}}{4}\mu_{B}g_{s}B_{0}$, and $\hbar g_{nc}=\mu_{B}g_{s}G_{nv}a_{0}$.
In the following part we assume $\Delta_{+}=\Delta_{-}=\Delta$ for simplicity, and define the bright and dark states as
\begin{equation}\begin{split}
\vert B\rangle&=\frac{1}{\sqrt{2}}(\vert +1\rangle+\vert -1\rangle),\\
\vert D\rangle&=\frac{1}{\sqrt{2}}(\vert +1\rangle-\vert -1\rangle).
\end{split}\end{equation}
It can be concluded that the state $\vert 0\rangle$ couples to the bright state $\vert B\rangle$, while the dark state $\vert D\rangle$ is decoupled. Then in the dressed state basis \{$\vert G\rangle=\cos\theta\vert 0\rangle-\sin\theta\vert B\rangle$, $\vert E\rangle=\sin\theta\vert 0\rangle+\cos\theta\vert B\rangle$\}, with $\tan 2\theta=2\sqrt{2}\Omega/\Delta$, the Hamiltonian $\hat{H}_{nc}$ including the free Hamiltonian $\hat{H}_{c}$ of the vibration mode can be rewritten as
\begin{equation}\begin{split}
\hat{H}_{nc}=&\hbar\omega_{c}\hat{a}_{c}^{\dag}\hat{a}_{c}+\hbar\omega_{eg}\vert E\rangle\langle E\vert+\hbar\omega_{dg}\vert D\rangle\langle D\vert\\
&+\hbar(g_{1}\vert G\rangle\langle D\vert+g_{2}\vert D\rangle\langle E\vert+H.c.)(\hat{a}_{c}+\hat{a}_{c}^{\dagger}),
\end{split}\end{equation}
where $\omega_{eg}=\sqrt{\Delta^{2}+8\Omega^{2}}$, $\omega_{dg}=(\Delta+\sqrt{\Delta^{2}+8\Omega^{2}})/2$, $g_{1}=-g_{nc}\sin\theta$, and $g_{2}=g_{nc}\cos\theta$.
Considering the condition $\Delta\gg\Omega$, we have $g_{1}\simeq 0$, $g_{2}\simeq g_{nc}$, $\omega_{eg}\simeq\Delta+4\Omega^{2}/\Delta$, $\omega_{dg}\simeq\Delta+2\Omega^{2}/\Delta$, and $\Lambda=\omega_{eg}-\omega_{dg}=2\Omega^{2}/\Delta$. Therefore, the $\hat{H}_{nc}$ under the RWA has the form
\begin{equation}
\hat{H}_{nc}/\hbar\approx\omega_{c}\hat{a}_{c}^{\dag}\hat{a}_{c}+\frac{1}{2}\Lambda\hat{\sigma}_{z}+g_{nc}(\hat{\sigma}_{+}\hat{a}_{c}+\hat{\sigma}_{-}\hat{a}_{c}^{\dagger}),
\end{equation}
with $\hat{\sigma}_{z}=\vert E\rangle\langle E\vert-\vert D\rangle\langle D\vert$, $\hat{\sigma}_{-}=\vert D\rangle\langle E\vert$, and $\hat{\sigma}_{+}=(\hat{\sigma}_{-})^{\dagger}$.

\providecommand{\noopsort}[1]{}\providecommand{\singleletter}[1]{#1}%

\end{document}